\let\Xdocument\document
\documentclass{iau}
\let\document\Xdocument

\usepackage{amsmath}
\usepackage{graphicx}
\usepackage{multirow}

\lefttitle{S.C. Tripathy et al. }
\righttitle{Probing the Sun's Near Surface Shear Layer}

\jnlPage{1}{6}
\jnlDoiYr{2024}
\doival{10.1017/xxxxx}

\aopheadtitle{Proceedings IAU Symposium}
\editors{A. Getling \&  L. Kitchatinov, eds.}

\title{Probing the Sun's Near Surface Shear Layer using HMI Spherical Harmonic Coefficients}

\author{{S.~C.~Tripathy$^1$}, {K.~Jain$^1$}, {S.~Kholikov$^{1,2}$} and {R.~Komm$^1$}}
\affiliation{$^1$ National Solar Observatory, 3665 Discovery Dr., Boulder, CO 80303, USA}
\affiliation{$^2$ Institute of Fundamental and Applied Research, National Research University, TIIAME, Uzbekistan}

\begin{document}
\begin{abstract}
We have measured zonal and meridional components of subsurface flows up to a depth of 30 Mm below the solar surface  by applying the technique of 
ring diagram on Dopplergrams which are constructed from the spherical harmonic (SH) coefficients.  The SH coefficients are obtained from the Helioseismic and Magnetic Imager 
(HMI) full-disk Dopplergrams. We find a good agreement and some differences between the flows obtained in this study 
with those from the traditional methods using direct Dopplergrams.  
 \end{abstract}

\begin{keywords}
Solar physics, Helioseismology, Solar interior, Solar oscillations, Solar activity, Solar convection zone, 
 Solar rotation, Solar meridional circulation
\end{keywords}

\maketitle
\vspace{-0.1in}
\section{Introduction} \label{sec:intro}

Helioseismic studies have illustrated that the most significant changes with the 
solar cycle occur in the near-surface shear layer (NSSL) occupying outer 5\% or
about 35 Mm of the Sun by radius.  Within this relatively thin layer, the density 
changes by several orders of magnitude and the rotation rate shows a local maxima. 
It is also believed that a nonlinear $\alpha\Omega$ dynamo could be operating in 
this layer where the velocity shear converts a part of the poloidal magnetic field 
into the toroidal field. The existence of this secondary dynamo is in addition to 
the global dynamo operating in the tachocline region and may be responsible for 
the small-scale field of the quiet Sun \citep{brandenburg13}. However, the dynamics 
of NSSL is poorly understood mainly because of the complex interactions between 
rotation, magnetism and convection. The advent of high-resolution Doppler 
observations from both space-borne and ground-based instruments, in conjunction 
with local helioseismic techniques such as ring-diagram \citep{hill88} and 
time-distance \citep{duvall93},  have made it possible to infer the subsurface 
flows in NSSL and their temporal evolution  
\citep{zhao04, basu_antia10, komm18, lin_chou18}. But local helioseismology techniques are limited in the depth range that can be probed. In particular  most of the studies using ring-diagram analysis is confined to a depth of 15~Mm. Only a handful of studies have used large-aperture tiles to analyse flows up to a depth of 30~Mm \citep{irene06, komm21}. 

In this context, we measure zonal and meridional components of the horizontal flows 
up to a depth of 30 Mm below the surface during the first six and half years of the
Helioseismic and Magnetic Imager (HMI) observations on board Solar Dynamics Observatory 
(SDO). We use a non-traditional method comprising of spherical harmonic (SH) coefficients
derived from the full-disk Dopplergrams and the ring-diagram (RD) technique. This method 
has the advantage to infer flows from low-resolution as well as high-resolution Dopplergrams.

\section{Data and Analysis Technique}
We  use HMI SH coefficient time series\footnote{series name: hmi.ap\_V\_sht\_gf\_72d} for $\ell$ = 0 to 400 obtained from the Joint 
Science Operation Center (JSOC) covering the period 2010 May 1 to 2016 December 31. 
Each time series is organized as series of individual ($n$, $\ell$, $m$) 
coefficients and are 72 days long.  We also download the window 
function  of each series to 
track the duty cycle.  Use of this SH time series has several
advantages as these have been generated from the velocity images which are 
corrected for known systematics, e.g., image scale, cubic distortion, 
CCD misalignment, inclination error and CCD tilt {\it etc.} The Dopplergrams 
were further detrended and gap filled. A description of all the corrections applied to 
the time series along with other details  are provided  in \cite{Larson18}.  

\begin{figure}
\center
\includegraphics[scale=.6]{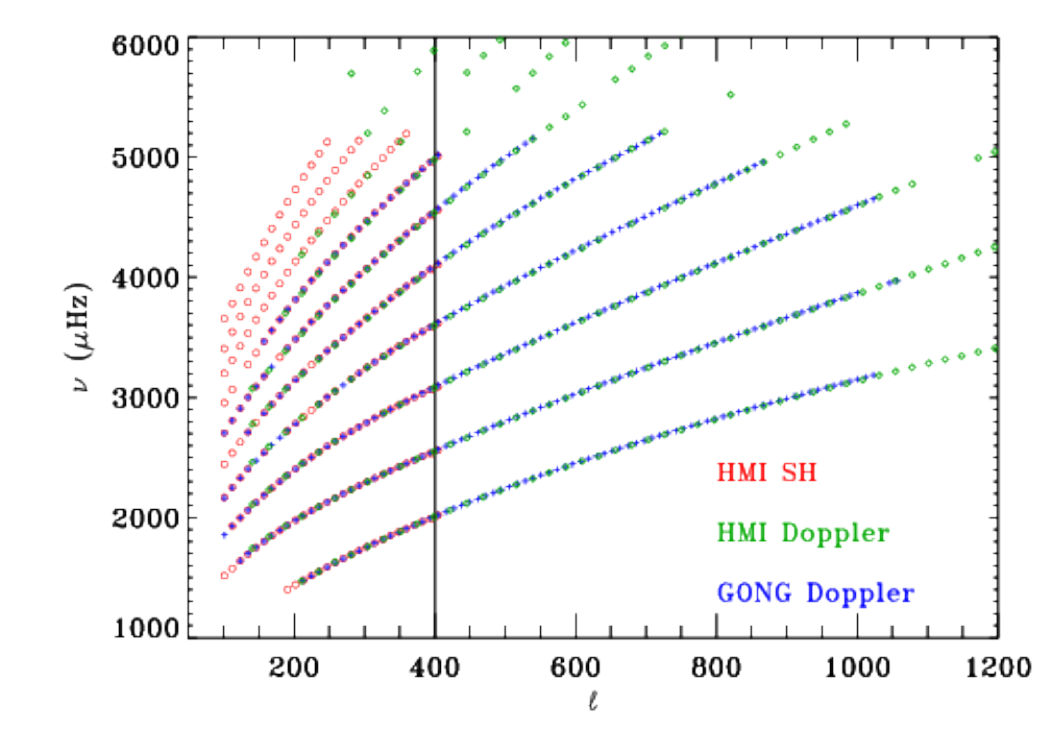}
\vskip -0.25cm
\caption{A typical $\ell$-$\nu$ diagram showing the distribution of the fitted modes 
at the disk center using HMI spherical harmonic coefficients (red circles), HMI Dopplergrams 
(green diamonds) and the GONG Dopplergrams (blue plus). Vertical line at $\ell$ = 400 
defines the boundary of the fitted modes using spherical harmonic coefficients.}
\label{lnu}
\end{figure}

Reconstruction of Dopplergrams using the SH coefficient time series (inverse 
SH decomposition) is  carried out following the procedure described in  
\citet{kholikov14_apj} 

\begin{equation}
V(\theta, \phi) = \sum_{\ell= 0}^{\ell = 400} \sum_{m =0}^{\ell} C_l^m P_l^m 
(\theta) e^{im\phi+\delta \phi} \; ,
\end{equation}

\noindent where $C_{\ell}^m$ are SH coefficients, $P_{\ell}^m$ is the associated 
Legendre polynomial of spherical degree $\ell$ and order $m, \theta$ and $\phi$ are 
latitude and longitude, respectively. We also use the differential rotation profile 
($\delta \phi$) of \citet{ libbrecht91} to remove the surface differential rotation 
where  tracking is carried out by introducing the surface differential rotation 
profile into the argument of the complex exponential in the right hand side of
Eq. (1). The reconstructed images were then remapped to 187 patches of the size
30$^\circ$ $\times$ 30$^\circ$ from disk center to $\pm$60$^\circ$ in latitude and 
$\pm 37.5^\circ$ in longitude spaced by  7.5$^\circ$. All the patches were tracked 
for 1664 min (the time series are divided in to 
1664 min time interval to produce the 
data cubes). A three-dimensional fast Fourier transform (FFT) is applied to 
each cube in both spatial and temporal directions to obtain a three-dimensional 
power spectrum which was subsequently processed through the RD 
 pipeline \citep{sct-corbard03}. The fitting algorithm in the RD pipeline 
uses the maximum-likelihood procedure described in \cite{anderson90} and fits 
about 270 modes for $n$-values between 0 -- 9 and $\ell$ values between 110 -- 400 
in the frequency range of 1400 -- 5200~$\mu$Hz. Figure~\ref{lnu} displays a typical 
$\ell$-$\nu$ diagram comparing  modes fitted at the disk center patch for three 
different data sets. 
It is evident that we 
recover most of the input modes in the  $\ell$ range of approximately 100 to 400 
(the RD technique does not fit modes below approximately $\ell$ of 100). We also note 
that we fit many more modes in the low $\ell$ and higher frequency ranges compared 
to methods using direct Dopplergrams. 

\begin{figure}
\center
\includegraphics[scale=.6]{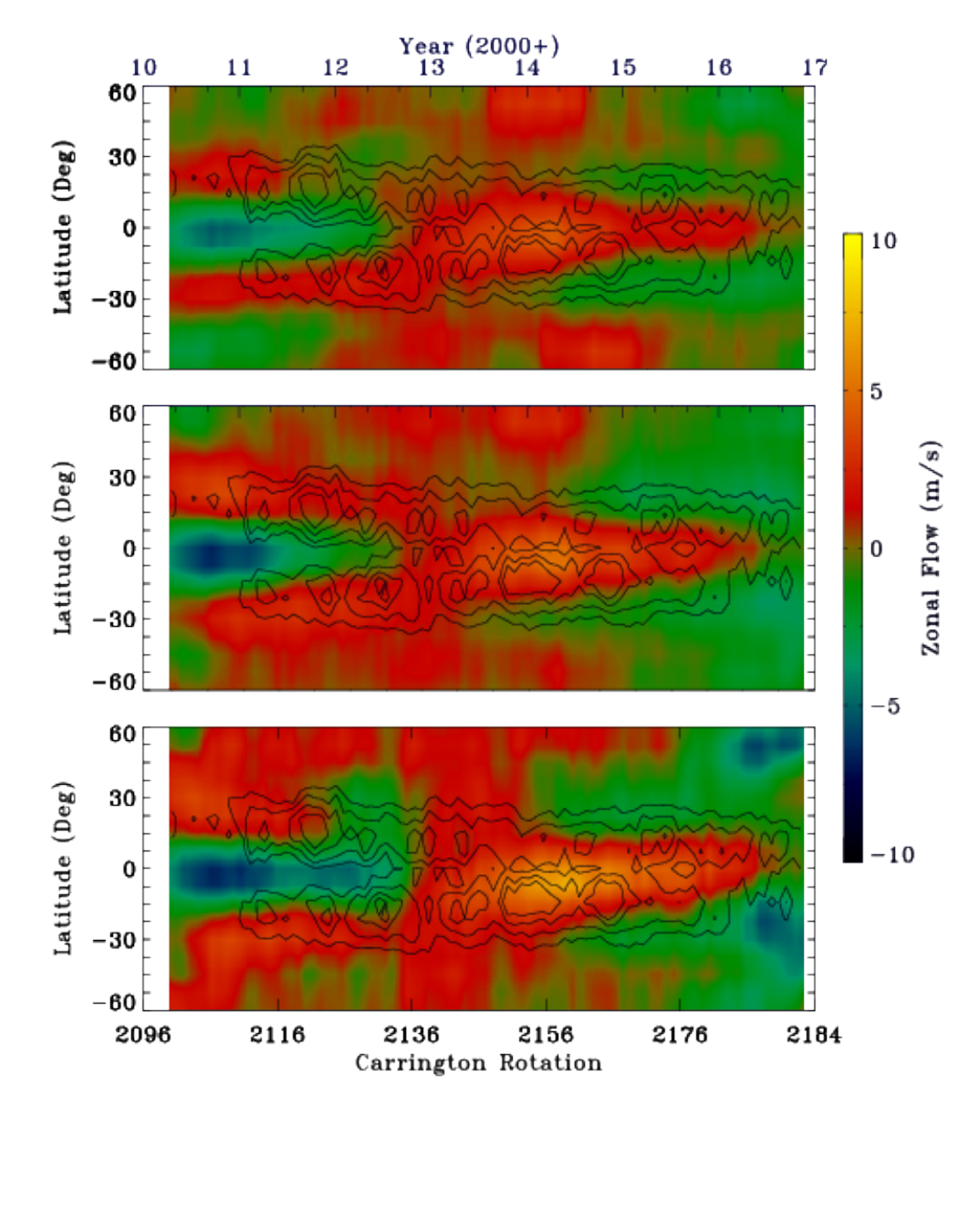}
\vskip -1.5cm
\caption{Temporal variation in zonal flow at three different depths: 5.2~$\pm$~2.2~Mm (top), 
15~$\pm$~7.2~Mm (middle) and 25.6~$\pm$ 7.2~Mm (bottom). The temporal mean has been subtracted at every 
latitude and depth. Black contours in each panel indicate the magnetic activity.}
\label{Sc_zonal}
\end{figure}

The subsurface zonal and meridional flows are finally derived using the Regularized 
Least Square (RLS) inversion method \citep{haber02}. The depth dependence in the 
inverted flows is known to have significant influence from the mode kernels used 
in the inversion. The kernels peak sharply near the surface owing to the availability 
of  significant number of modes but are less localized with increasing 
depth. Thus the depth resolution turns out to be poor resulting in 
relatively larger errors in deeper layers \citep{Jain17}. However, in this study 
we have a limited set of high-degree modes and thus we find that the errors 
associated with the flows measured closer to the surface are larger compared to 
deeper depths. We also correct the flows for the annual variation with the solar 
inclination angle towards Earth (B$_0$ angle) and other systematic effects following 
the procedure described in \citet{Komm15}.

\section{Results}
Figure~\ref{Sc_zonal} shows the zonal flow at three depths (5.2 $\pm$~2.2~Mm, 
15~$\pm$~7.2~Mm, and 25.6~$\pm$~7.2~Mm) as a function
of time and latitude where the mean is subtracted at each latitude and depth. The contours 
on top of the flows are plotted to depict the evolution of magnetic activity. The figure 
illustrates the bands of slower and faster rotation relative to the mean flow, known as 
torsional oscillations, of few m/s at all depths. These are comparable with  those 
obtained in the traditional methods \citep[e.g.,][for ring-diagram and time-distance 
analyses, respectively]{komm18,zhao14}. As is well known, we find that the bands of 
faster rotation at mid latitudes are well aligned with the migration of magnetic activity 
towards the equator but appears about a year earlier before the activity appears at 
the surface. Although the signature of the poleward branch is visible starting around 
2012 during the ascending phase of the cycle, we do not clearly see the branch probably 
due to the weak cycle 24. Similar results are reported in torsional oscillations derived 
from the global modes 
\citep{howe2018}. Also, the connection between the mid-latitude branch in
this solar cycle with the polar branch of the previous cycle  is not apparent 
in the figure due
to the temporal coverage. Nevertheless, a close comparison of the zonal flows obtained in 
\citet{komm18} displays similarities as well as some differences in finer details. 
These differences  probably  arise due to the different patch sizes used in these 
studies covering different areas at each grid point and the different smoothing parameters.
However the overall patterns are similar in both the studies. Further, we find a 
discontinuity in the mid-latitude branch in the Northern hemisphere around 2012 in the 
top and bottom panels of Figure~\ref{Sc_zonal} corresponding to 5 and 25 Mm (the band 
is barely visible in 5~Mm). While the traditional methods provide
continuous branches in both hemispheres, our analysis shows the continuous branch to be present in the stronger hemisphere only. It is not clear if this effect is  due to smaller temporal coverage which provides different mean values over the two hemispheres. It is also plausible that the corrections that is applied at deeper depths to remove the systematic variations are not optimized. 

\begin{figure}
\center
\includegraphics[scale=.6]{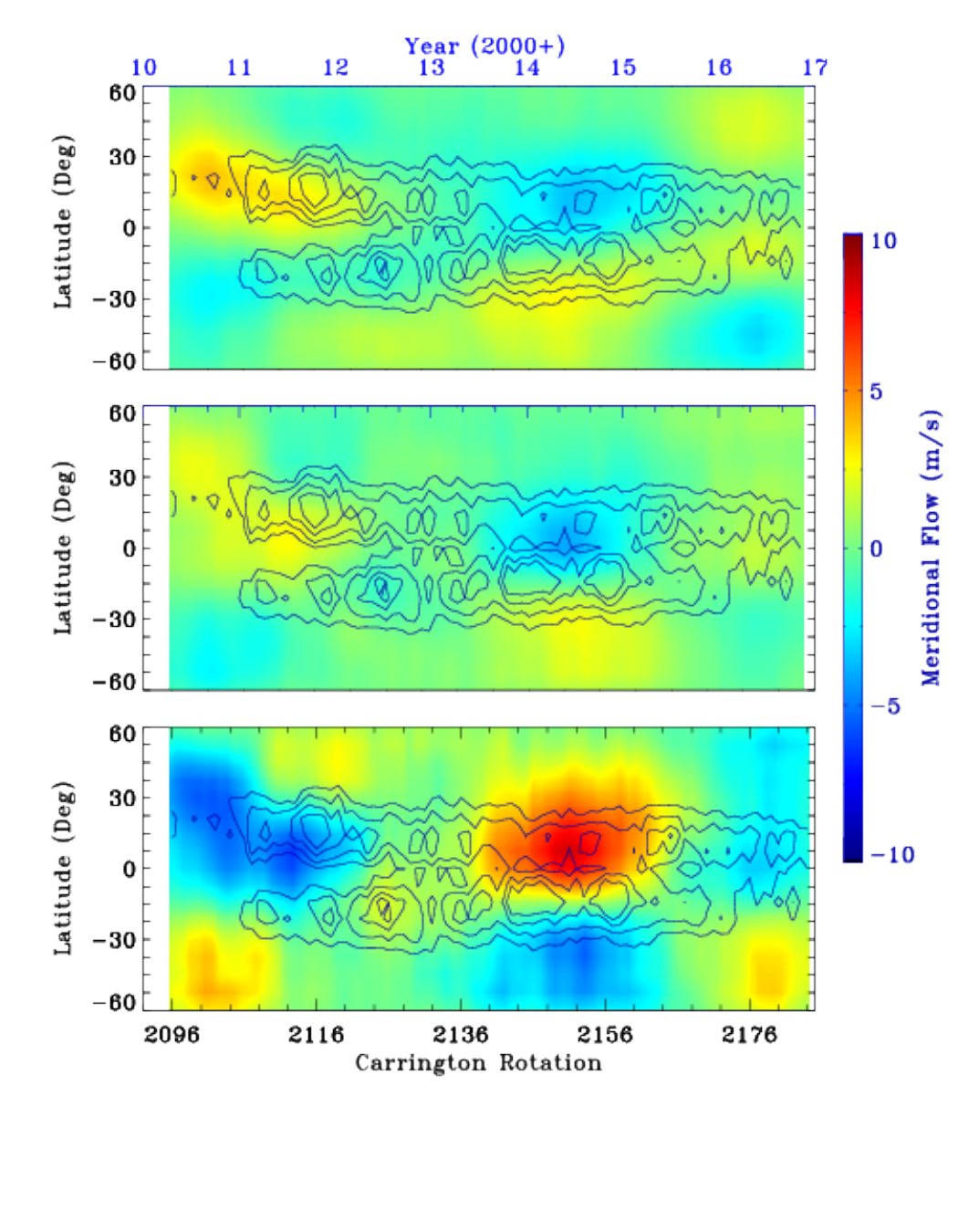}
\vskip -1.5cm
\caption{Same as Figure~2 but for meridional flows}
\label{Sc_meri}
\end{figure}

\begin{figure}
\center
\includegraphics[scale=.6]{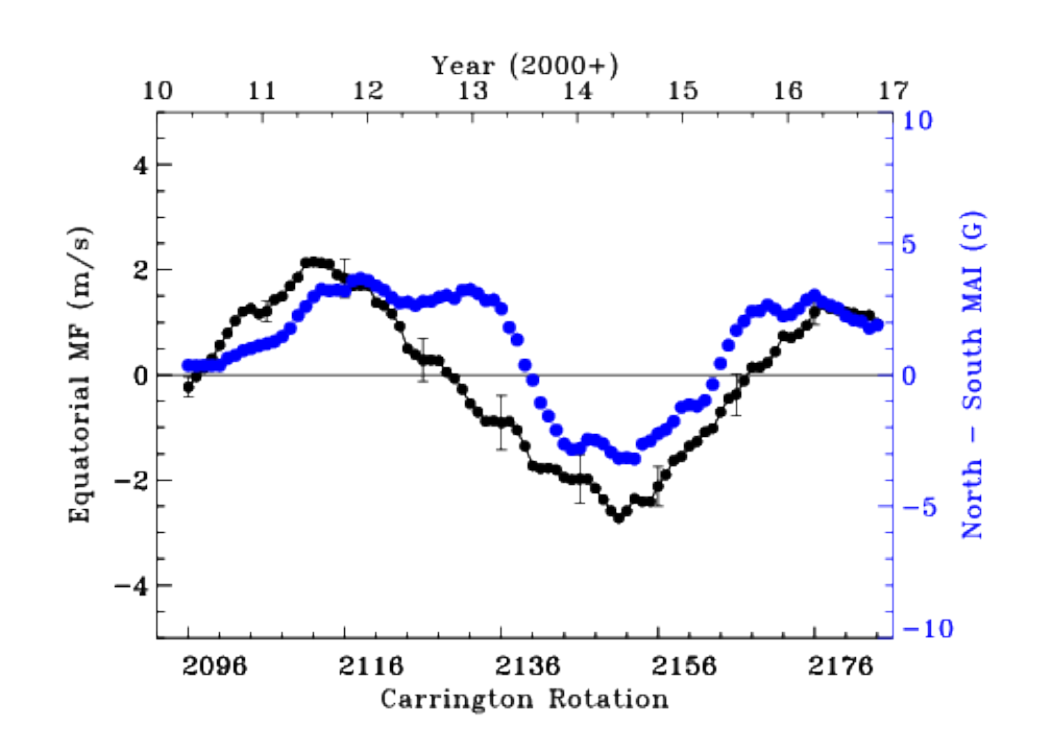}
\caption{The equatorial meridional flow averaged over the depth range of 
0.9 -- 3.4 Mm. Positive values indicate northward flows, while negative values 
indicate southward flows. The blue line represents  the strength of the magnetic 
activity where the MAI of 7.5$^\circ$~S tile is subtracted from the 7.5$^\circ$~N tile.  For clarity, the error bars for flow measurements are shown only for few points.}
\label{cross}
\end{figure}

It is well recognized that the meridional flow is poleward on the surface and in 
shallow depths and increases gradually with depth. We also find similar results
with a peak amplitude of about 15~m/s near the surface and 30~m/s at a depth of 29~Mm.  Figure~\ref{Sc_meri} displays the temporal 
evolution of the meridional flow  relative to the mean flow at
each latitude and depth for the same depths as the zonal flows.   We obtain bands of faster and slower meridional circulation residuals with time 
but with opposite signs in both hemispheres implying poleward flows as positive
values indicate flows to the north and negative values for the flows to the south. At the 
shallow depths, the band of poleward meridional flow appears at about 60$^\circ$ 
latitude around the mid-2010 (beginning of our data) and slowly migrates towards the equator 
reaching there in about 2013.   A similar result is reported in \citet{komm18}. 
But at deeper depths, the flow pattern is reversed; the fast band has changed to the slow band and vice versa and differs from the result of \cite{komm21} where results from HMI 30$^\circ$ tiles were used. However, we should note that the HMI RD result has a coarser resolution in latitude and longitude since the flow measurement are available only at a grid spacing of 15$^\circ$.  

The amplitude and direction of the meridional flow at the equator is an important
topic in the context of cross-equatorial cancellation of magnetic flux and 
formation of polar flux. Early investigations that found flows crossing the 
equator believed it to be an error in the alignment of  the telescope since its 
magnitude is one order smaller than the maximum amplitude of the meridional flow. 
But recent advances in the surface flux transport models indicate that it plays 
a crucial role in transporting magnetic flux across the equator from the 
dominant hemisphere to the other one and thus cancels the excess flux. We
further investigate, as shown in Figure~\ref{cross}, the meridional flows at 
the equator at a depth closer to the surface 
(averaged over 0.9 -- 3.4 Mm). It is evident 
that the meridional flows at the equator are small but non-zero. It is mainly 
positive at the beginning of cycle and oscillates between positive and negative 
values as the solar cycle progresses which is believed to be a consequence 
of the imbalance of activity between the two hemispheres. To quantify this flux 
imbalance,  we subtract the magnetic activity index (MAI) of 7.5$^\circ$~S tiles from 7.5$^\circ$~N tiles.  The MAI values for each tile is calculated by tracking and re-mapping 
line-of-sight magnetograms  in the same fashion as Dopplergrams for the same length of time.
The absolute values of all pixels higher than a defined threshold are then averaged 
to compute the unsigned proxy for magnetic field (MAI).
It is evident from Figure~\ref{cross} that 
the cross equatorial flows follow the hemisphere with dominant activity. The 
Pearson's linear cross correlation is found to be 75\% confirming the relation 
between the cross-equatorial flow and strength of the magnetic activity. These
findings are in agreement with those reported in \citet{komm22} where subsurface 
flow measurements were derived from 15$^\circ$ tiles.     

\section{Summary}
Reconstructing Dopplergrams from the HMI SH coefficients and subsequently applying 
the ring-diagram technique, we have derived subsurface flows up to a depth of 30 Mm 
from mid-2010 to the end of 2016. Our preliminary results of zonal and meridional 
flow profiles mostly agree near the surface and intermediate depth ranges with 
other studies using observed Dopplergrams. However, the measured meridional flows 
at greater depths show a flow pattern which is reversed compared to the flows 
observed near the surface. Although there are limited helioseismic studies which 
probes the flows at depths greater than 15 Mm using RD technique, our result appears to differ from  
these findings. It is plausible that the applied empirical correction for the systematic variations has not removed all artifacts. 
We plan to revisit the correction scheme when flow measurements for whole cycle 24 is computed.    
 
The data used here are courtesy of NASA/SDO and the HMI Science Team. This work 
was supported by NASA grants 80NSSC20K0194, 80NSSC21K0735, and 80NSSC23K0404  to 
the National Solar Observatory. SCT, KJ and RK acknowledge the partial financial 
support from NASA Cooperative Agreement 80NSSC22M0162 to Stanford University for 
the COFFIES Drive Science Center.


\begin{thebibliography}{}
\bibitem[Anderson, Duvall, and Jefferies(1990)]{anderson90}Anderson, E.R., Duvall, T.L., and Jefferies, S.M.: 1990, {\it ApJ}, 364, 699

\bibitem[{{Basu} \& {Antia}(2010)}]{basu_antia10}{Basu}, S., \& {Antia}, H.~M. 2010, {\it ApJ}, 717, 488
\bibitem[Brandenburg(2013)]{brandenburg13}Brandenburg, A.: 2013, {\it Solar and Astrophysical Dynamos and Magnetic Activity} 294, 387
\bibitem[{{Corbard} {et~al.}(2003){Corbard}, {Toner}, {Hill}, {Hanna}, {Haber},  {Hindman}, \& {Bogart}}]{sct-corbard03}{Corbard}, T., {Toner}, C., {Hill}, F., {et~al.} 2003, in ESA Special
  Publication, Vol. 517, GONG+ 2002. Local and Global Helioseismology: the  Present and Future, ed. H.~{Sawaya-Lacoste}, 255

\bibitem[{{Duvall} {et~al.}(1993){Duvall}, {Jefferies}, {Harvey}, \&  {Pomerantz}}]{duvall93}
{Duvall}, T.~L., J., {Jefferies}, S.~M., {Harvey}, J.~W., \& {Pomerantz}, M.~A.  1993, {\it Nature}, 362, 430

\bibitem[Gonz{\'a}lez Hern{\'a}ndez \emph{et al.}(2006)]{irene06}Gonz{\'a}lez Hern{\'a}ndez, I., Komm, R., Hill, F., Howe, R., Corbard, T., and Haber, D.A.: 2006, {\it ApJ} 638, 576 

\bibitem[{{Haber} {et~al.}(2002){Haber}, {Hindman}, {Toomre}, {Bogart},  {Larsen}, \& {Hill}}]{haber02}
{Haber}, D.~A., {Hindman}, B.~W., {Toomre}, J., {et~al.} 2002, {\it ApJ}, 570, 855

\bibitem[{{Hill}(1988)}]{hill88}{Hill}, F. 1988, {\it ApJ}, 333, 996

\bibitem[Howe {et~al.}(2018)]{howe2018}{Howe}, R., {Hill}, F., {Komm}, R., {Chaplin}, W.~J., {Elsworth}, Y., {Davies}, G.~R., and, ...: 2018, {\it ApJ}, 862, L5 

\bibitem[{{Jain} {et~al.}(2017){Jain}, {Tripathy}, \& {Hill}}]{Jain17}
{Jain}, K., {Tripathy}, S.~C., \& {Hill}, F.\ 2017, {\it ApJ}, 849, 94

\bibitem[{{Kholikov} {et~al.}(2014){Kholikov}, {Serebryanskiy}, \&  {Jackiewicz}}]{kholikov14_apj}
{Kholikov}, S., {Serebryanskiy}, A., \& {Jackiewicz}, J. 2014, {\it ApJ}, 784, 145
\bibitem[Komm(2021)]{komm21}Komm, R.: 2021, {\it Solar Physics} 296, 174

\bibitem[Komm(2022)]{komm22}Komm, R.: 2022, {\it Solar Physics}, 297, 99

\bibitem[{{Komm} {et~al.}(2015){Komm},  {Gonz{\'a}lez Hern{\'a}ndez}, {Howe}, \& {Hill}}]{Komm15}{Komm}, R., {Howe}, R.,  {Gonz{\'a}lez Hern{\'a}ndez}, I.,  \& {Hill}, F. 2015, {\it Solar Physics}, 290, 1081


\bibitem[{{Komm} {et~al.}(2018){Komm}, {Howe}, \& {Hill}}]{komm18}{Komm}, R., {Howe}, R., \& {Hill}, F. 2018, {\it Solar Physics}, 293, 145

\bibitem[{{Larson} \& {Schou}(2018)}]{Larson18}{Larson}, T.~P., \& {Schou}, J. 2018, {\it Solar Physics}, 293, 29

\bibitem[Libbrecht and Morrow(1991)]{libbrecht91}Libbrecht, K.G. and Morrow, C.A.: 1991, {\it Solar Interior and Atmosphere}, 479

\bibitem[{{Lin} \& {Chou}(2018)}]{lin_chou18}{Lin}, C.-H., \& {Chou}, D.-Y. 2018, {\it ApJ}, 860, 48

\bibitem[{{Zhao} \& {Kosovichev}(2004)}]{zhao04}{Zhao}, J., \& {Kosovichev}, A.~G. \ 2004, {\it ApJ}, 603, 776

\bibitem[{{Zhao} {et~al.}(2014) {Zhao}, {Kosovichev}, \&  {Bogart}}]{zhao14}  {Zhao}, J., \& {Kosovichev}, A.~G. \ 2014, {\it ApJ}, 789, L7

\end{thebibliography}
\end{document}